\documentclass{aastex63}
\usepackage{mathrsfs}
\usepackage{threeparttable}

\begin{document}

\title{Chasing the neutrino blazar candidates II: SED modeling with hadronic model}

\correspondingauthor{Hubing Xiao}
\email{hubing.xiao@shnu.edu.cn}
\correspondingauthor{Junhui Fan}
\email{fjh@gzhu.edu.cn}

\author[0000-0001-8244-1229]{Hubing Xiao}
\affiliation{Shanghai Key Lab for Astrophysics, Shanghai Normal University, Shanghai 200234, People’s Republic of China}

\author[0009-0000-4102-9115]{Zhihao Ouyang}
\affiliation{Shanghai Key Lab for Astrophysics, Shanghai Normal University, Shanghai 200234, People’s Republic of China}

\author[0000-0001-7416-7434]{Lili Yang}
\affiliation{School of Physics and Astronomy, Sun Yat-sen University, Zhuhai 519082, People’s Republic of China}

\author[0000-0002-8206-5080]{Jingtian Zhu}
\affiliation{School of Network Engineering, Wuhu Vocational Technical University, Anhui 241000, People’s Republic of China}

\author[0000-0002-4455-6946]{Minfeng Gu}
\affiliation{Shanghai Astronomical Observatory, Chinese Academy of Sciences, Shanghai 200030, People’s Republic of China}

\author[0000-0002-1908-0536]{Liang Chen}
\affiliation{Shanghai Astronomical Observatory, Chinese Academy of Sciences, Shanghai 200030, People’s Republic of China}

\author[0000-0001-8485-2814]{Shaohua Zhang}
\affiliation{Shanghai Key Lab for Astrophysics, Shanghai Normal University, Shanghai 200234, People’s Republic of China}

\author[0009-0009-1617-8747]{Zhijian Luo}
\affiliation{Shanghai Key Lab for Astrophysics, Shanghai Normal University, Shanghai 200234, People’s Republic of China}

\author[0000-0002-5929-0968]{Junhui Fan}
\affiliation{Center for Astrophysics, Guangzhou University, Guangzhou 510006, People’s Republic of China}
\affiliation{Great Bay Brand Center of the National Astronomical Data Center, Guangzhou 510006, People’s Republic of China}
\affiliation{Key Laboratory for Astronomical Observation and Technology of Guangzhou, Guangzhou 510006, People’s Republic of China}
\affiliation{Astronomy Science and Technology Research Laboratory of Department of Education of Guangdong Province, Guangzhou 510006, People’s Republic of China}

\begin{abstract}
Blazars are promising candidates for high-energy neutrino sources, yet the physical origin of their neutrino emission remains uncertain. 
In this work, we extend our previous study by modeling the broadband spectral energy distributions (SEDs) of 103 neutrino blazar candidates (NBCs) within a hadronic framework. 
To estimate the maximum possible neutrino output, we adopt an assumption in which the high-energy emission is dominated by $p\gamma$ interactions and the contribution from leptonic inverse Compton scattering is strongly suppressed.
From the SED modeling, we constrain nine key parameters describing the emission region and particle energy distributions. 
We perform a partial correlation analysis to investigate the relationship between neutrino luminosity and electromagnetic emission, and we found a weak/moderate correlation between optical $R$ band and neutrino emission.
Our model predicts prominent proton synchrotron emission peaking in the MeV band for most sources, with 99 out of 103 NBCs exhibiting proton synchrotron peaks within 0.1–100 MeV, highlighting the MeV band as a key window for distinguishing between leptonic and hadronic scenarios. 
Based on the model-predicted maximum neutrino fluxes, we find that three NBCs are potentially detectable by IceCube, while up to 22, 45, and 62 sources may be detectable by KM3NeT, NEON, and TRIDENT, respectively. 
These results provide testable predictions for future multi-messenger observations and offer new insights into the composition and radiation mechanisms of blazar jets.

\end{abstract}


\section{Introduction}
\label{sec:intro}
Blazars, the most extreme subclass of active galactic nuclei (AGNs), exhibit a range of remarkable observational properties, including rapid and strong variability, high and variable polarization, intense and variable $\gamma$-ray emission, and apparent superluminal motion \citep{Wills1992, Urry1995, Fan2002, Fan2004, Rani2013, Fan2014, Lyutikov2017, Xiao2019, Xiao2022MNRAS}.
Based on their optical spectral characteristics, blazars are commonly divided into two subclasses: BL Lacertae objects (BL Lacs) and flat-spectrum radio quasars (FSRQs).
BL Lacs are characterized by weak or absent emission lines, with rest-frame equivalent widths $\mathrm{EW} < 5~\mathrm{\AA}$, whereas FSRQs exhibit prominent broad emission lines with $\mathrm{EW} \geq 5~\mathrm{\AA}$ \citep{Urry1995, Scarpa1997}.
Blazar emission is characterized by a two-bump broadband spectral energy distribution (SED) in the $\log \nu$–$\log \nu F_{\nu}$ representation.
The low-energy component, peaking from the millimeter to soft X-ray bands, is generally attributed to synchrotron radiation from relativistic electrons.
In contrast, the origin of the high-energy component, which typically peaks in the MeV–GeV range, remains debated and is commonly interpreted within either leptonic or hadronic emission scenarios.

The development of leptonic models has been well established over several decades.
In the 1960s, it was recognized that the nonthermal emission from radio sources, spanning from the radio to optical bands, can be attributed to synchrotron radiation from relativistic electrons \citep{Westfold1959ApJ130, Ginzburg1965ARA&A3, Kellermann1966ApJ146}.
It was subsequently realized that these same electrons can produce high-energy photons through inverse Compton (IC) scattering off ambient soft photon fields, e.g., the synchrotron photons \citep{Jones1974ApJ188, Ghisellini1985A&A146}, the broad-line region (BLR) \citep{Sikora1994, Fan2006}, or the dusty torus (DT) \citep{Blazejowski2000, Arbeiter2002, Sokolov2005}.
The leptonic scenario was further consolidated following the detection of $\gamma$-ray emission by the Energetic Gamma Ray Experiment Telescope (EGRET) \citep{Maraschi1992ApJ397}, and it has since become the most widely accepted framework for modeling blazar SEDs after the launch of the \textit{Fermi} Large Area Telescope (\textit{Fermi}-LAT).
Given that cosmic rays are predominantly composed of protons, it was natural to consider whether protons might also play a significant role in blazar jets, thereby motivating the development of hadronic models for blazar emission.
In early 1990s, \citep{Mannheim1993A&A269} proposed protons within the jet are accelerated to ultra-high energies and subsequently interact with ambient photon fields through photo-meson processes ($p-\gamma$ interactions) to produce secondary particles, which in turn emit $\gamma$-rays.
\citet{Aharonian2000NewA5} employed hadronic models to explain the TeV emission from BL Lac objects, although such models often require jet powers that exceed the Eddington limit \citep{Bottcher2013}.
The detection of high-energy neutrino emission has provided the first multi-messenger evidence for the presence of relativistic hadrons in blazar jets \citep{IceCube2018Sci_1}.
The hadronic models have gained more attention for producing neutrinos through charged pion-chain decay, and this mechanism is naturally applied to explain the neutrino detection of TXS 0506+056 and to make predictions for other sources \citep{Cerruti2019MNRAS483, Xue2019ApJ886, Xue2021ApJ906, Wang2024ApJS271, Ouyang2025ApJ980}.

Neutrino astronomy is widely regarded as a promising avenue for addressing the century-old questions of the origin of cosmic rays and the mechanisms responsible for their acceleration.
In the context of blazar studies, neutrinos serve as powerful probes of jet composition, the physical environment near supermassive black holes, and the contribution of blazars to the diffuse high-energy neutrino background.
These fundamental questions can, in principle, be investigated through blazars that emit neutrinos; however, at present, TXS~0506+056 remains the only blazar with a firmly established neutrino association \citep{IceCube2018Sci_1}.

In our previous work \citep[][hereafter Paper~I]{Zhu2024ApJS275}, we employed a few-shot learning approach to identify neutrino blazar candidates (NBCs) from \textit{Fermi}-LAT detected blazars \citep{Abdollahi2022ApJS260}, resulting in a sample of 199 NBCs.
In this work, with the aim of extending our previous study and investigating the physical properties of NBCs, we adopt a hadronic model to construct broadband SEDs for these sources.
Similar modeling frameworks have been adopted in previous studies \citep[e.g.,][]{Cerruti2015MNRAS, Abe2025MNRAS540}, although several modifications are introduced in the present work.
In order to estimate the maximum possible neutrino flux, we assume that all high-energy (HE) and very-high-energy (VHE) photons are produced through $p\gamma$ interactions and the associated electromagnetic cascades.
The paper is arranged as follows: 
in Section 2, we introduce the SED fitting method;
in Section 3, we present the results;
in Section 4, we present the discussions;
the conclusion is presented in Section 5.

\section{Sample and Method}
\label{sec:sam&mtd}
We searched for multi-wavelength observations for the 199 neutrino blazar candidates (NBCs; see Paper~I) by matching the associated name provided by the \textit{Fermi}-LAT Fourth Source Catalog \citep[4FGL,][]{Abdollahi2022ApJS260} with items in Firmamento database\footnote{\url{https://firmamento.nyuad.nyu.edu/data_access}}
 \citep{Tripathi2024AJ167}.
During the data collection process, we identified 103 NBCs with sufficient multi-wavelength coverage that exhibit a clear, characteristic two-hump blazar SED structure by visual inspection. 
These data span from the radio to the $\gamma$-ray bands and were obtained from different observational campaigns.
In addition, TXS~0506+056 is included in this work, and its multi-wavelength data are used to construct the SED and to compare its properties with those of the NBCs.
Typically, there are two approaches can be adopted when handling multi-wavelength data prior to broadband SED construction.
One approach is to select quasi-simultaneous data within a narrow time window, which is suitable for constraining emission region properties and testing specific SED models \citep[e.g.,][]{Tan2020, Wang2024ApJS271}. 
Alternatively, unfiltered archival data can be used to investigate the general properties of a blazar population \citep[e.g.,][]{Xiao2025ApJ991}. 
In this work, we adopt the latter approach and use the multi-wavelength data without temporal filtering.
Consequently, both quiescent and flaring states are included in the broadband SEDs constructed for the 103 sources, enabling us to investigate their general, population-level properties rather than state-dependent behaviors.
Motivated by the current understanding that blazar jets are composed of both leptons and hadrons, we employ a hadronic model to construct the broadband SEDs in this study.

The one-zone hadronic model is employed to reproduce the broadband emissions and neutrino emissions using the self-consistent, time-dependent numerical code AM$^{3}$ \citep{Gao2017ApJ843, Klinger2024ApJS275}.
We assume a spherical emitting blob/region of radius $R_{\rm em}$, immersed in a homogeneous magnetic field of strength $B$, moving with a bulk Lorentz factor $\Gamma$ along the relativistic jet. 
This results in a Doppler factor of $\delta = \frac{1}{\Gamma (1 - \beta \cos \theta)}$, where $\beta c$ is the velocity of the emission blob, $c$ is the speed of light, and $\theta$ is the viewing angle of the jet with respect to our line of sight. 
Given that the viewing angle is less than a few degrees for blazars ($\theta \leq 1/\Gamma$), $\Gamma \simeq \delta$ can be obtained. 
We assume the non-thermal electrons/protons are accelerated and injected into the emission blob with an injection rate following a power-law spectrum $\frac{\mathrm{d}^{2} N_{\rm e \, (p)}}{\mathrm{d} \gamma_{\rm e (p)} \, \mathrm{d} t} \propto \gamma^{- \alpha_{\rm e \, (p)}}$ ranging from $\gamma_{\rm e (p)}^{\rm min}$ up to $\gamma_{\rm e (p)}^{\rm max}$, where the $\gamma_{\rm e(p)}$ and $\alpha_{\rm e \, (p)}$ are the Lorentz factor and the spectral index of the electrons/protons, respectively\footnote{Quantities with the superscript ``obs'' refer to the observed frame, those with ``rest'' refer to the rest frame of the AGN, and quantities without a superscript are defined in the rest frame of the relativistic jet.}. 
The injected power from electrons/protons is characterized by the respective injection luminosity $L_{\rm e (p), \, inj}$.
Then, the time-dependent transport equation is solved to obtain the steady-state particle distributions:
\begin{equation}
    \frac{\partial N(\gamma, \, t)}{\partial t} = Q(\gamma, \, t) - \frac{\partial}{\partial \gamma} \left(\dot{\gamma}(\gamma, \, t)N(\gamma, \, t)  \right) - \frac{N(\gamma, \, t)}{t_{\rm esc}} ,
\end{equation}
where $Q(\gamma , t)$ denotes the particle injection rate, $\dot{\gamma}(\gamma , t)$ is the energy loss rate, and $t_{\rm esc}$ is the escape timescale of particles.
The escape timescale is assumed to be comparable to the dynamical timescale $t_{\rm esc} \sim t_{\rm dyn} = R/c$, which is widely adopted in the literature \citep[e.g, ][]{Wang2024ApJS271, Prince2024MNRAS527, Rodrigues2024A&A681A, Jiang2025ApJ986}.
A steady-state is achieved when the injection is balanced by cooling and/or escape within the emission region.
Then, in the steady-state, the electrons produce synchrotron radiation and contribute to the low-energy hump after synchrotron self-absorption, 
the protons interact with the photons produced in the synchrotron radiation and contribute to the high-energy hump through the photo-pion production ($p\gamma \rightarrow n \pi^{+}$ or $p \pi^{0}$) and Bethe-Heitler pair production ($p\gamma \rightarrow p \, e^{+} e^{-}$).
Moreover, the high-energy photons can undergo photon-photon pair production and annihilation with the low-energy photons ($\gamma \gamma \rightarrow e^{+} e^{-}$). 
Neutrinos are produced through the $\pi^{+}$ decay ($\pi^{+} \rightarrow \mu^+ \nu_{\mu}$) and the subsequent $\mu^{+}$ decay ($\mu^+ \rightarrow e^+ \nu_e \bar{\nu}_{\mu}$) from the photo-pion production. 
The secondary electron–positron pairs produced by these processes, including photo-pion production, Bethe–Heitler pair production, and photon–photon annihilation, can subsequently radiate via synchrotron and inverse Compton processes.
In order to estimate the neutrino emission as an upper limit from purely hadronic origins, we consider the high-energy hump is mainly contributed from the hadronic processes and suppress the IC scattering from leptons.
At very high energies, gamma-ray photons interact with the extragalactic background light (EBL), resulting in significant attenuation of the observed flux.
This effect is included in our modeling using the EBL model from \citet{Dominguez2011}, applied after modeling for each source.
In the present work, the low-energy component of the SED, spanning from the radio to the optical/ultraviolet or X-ray bands, is attributed to synchrotron radiation from relativistic electrons.
At higher energies, the X-ray emission is interpreted as synchrotron radiation from either relativistic electrons or protons, while the GeV $\gamma$-ray emission is attributed to hadronic interactions, predominantly via Bethe–Heitler pair production.

For some sources, we employ blackbody thermal function to fit an additional bump appeared in the frequency range of 10$^{14}$ to 10$^{15}$ Hz, and characterized it using peak frequency $\nu_{\rm thermal}^{\rm rest}$ and thermal luminosity $L_{\rm thermal}^{\rm rest}$.

Based on the physical setup described above, the one-zone hadronic model is characterized by 11 parameters; for sources with an additional thermal hump, two more parameters are included, bringing the total to 13 parameters.
The model parameter spaces for all parameters are listed in Table \ref{tab:params_space}. 
In particular, we fix the minimum Lorentz factor of protons to $\gamma_{\rm p}^{\rm min} = 10^2$, and the spectral index of proton spectrum to $\alpha_{\rm p} = 1.0$.
During the fitting process, radio data below $10^{11}$ Hz are excluded from our fitting, as the one-zone model cannot adequately reproduce the observed radio emission due to synchrotron self-absorption. 
We perform the SED fitting of the multi-wavelength data using the PySwarms\footnote{\url{https://pyswarms.readthedocs.io/en/latest/index.html}} package, which implements the Particle Swarm Optimization (PSO) algorithm \citep{MirandaJOSS2018pyswarms}.
This method simulates a population of particles that explore the parameter space collectively to minimize the objective function and is well-suited for exploring high-dimensional parameter spaces.
We note, however, that the PSO method may not always converge to the global optimum, but can instead settle in a local minimum, potentially leading to suboptimal SED representations. 
To mitigate this effect, an additional ``fit-by-eye'' procedure is employed to refine the parameters and obtain an acceptable representation of the SED.

\begin{table}[htbp]
\centering
\caption{The model parameter space}
\label{tab:params_space}
\begin{tabular}{llc}
\hline
Parameter                   				&   Space                 	& Units			\\
\hline
$\log R_{\rm em}$          				&   (14, 17)               	&   cm  			\\
$\log B$                          				&   (-2, 2)                  	&   G  			\\
$\Gamma$                     				&   (1, 30)                  &   --  			\\
$\log L_{\rm e}$    					&   (38, 45)    		&   erg s$^{-1}$  	\\
$\log \gamma_{\rm e}^{\rm min}$    		&   (1, 5)    		&   --  			\\
$\log \gamma_{\rm e}^{\rm max}$    		&   (3, 6)    		&   --  			\\
$\alpha_{\rm e}$    					&   (1, 3.5)    		&   --  			\\
$\log L_{\rm p}$    					&   (42, 47)    		&   erg s$^{-1}$  	\\
$\log \gamma_{\rm p}^{\rm min}$    		&   Fixed to 2.0    	&   --  			\\
$\log \gamma_{\rm p}^{\rm max}$    		&   (5, 9)    		&   --  			\\
$\alpha_{\rm p}$    					&   Fixed to 1.0    	&   --  			\\
\hline
$\log \nu^{\rm rest}_{\rm thermal}$		&   (14, 15)    		&   Hz  			\\
$\log L_{\rm thermal}^{\rm rest}$    		&   (40, 48)   		&   erg s$^{-1}$  	\\ 
\hline
\end{tabular}
\end{table}

\section{Results}
\label{sec:dis}
\subsection{The results of SED fitting}
\begin{figure}
\centering
\includegraphics[width=250pt]{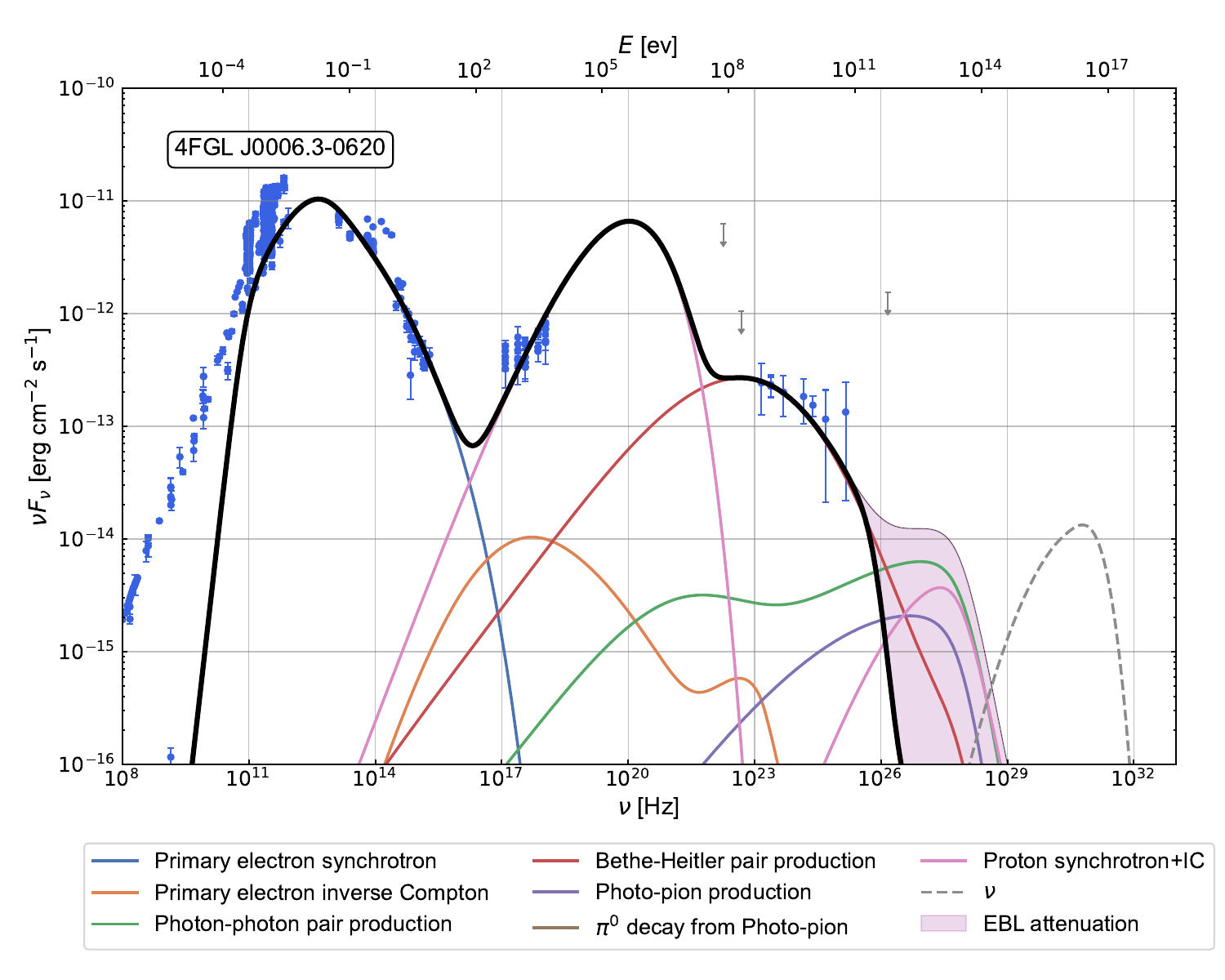}
\includegraphics[width=250pt]{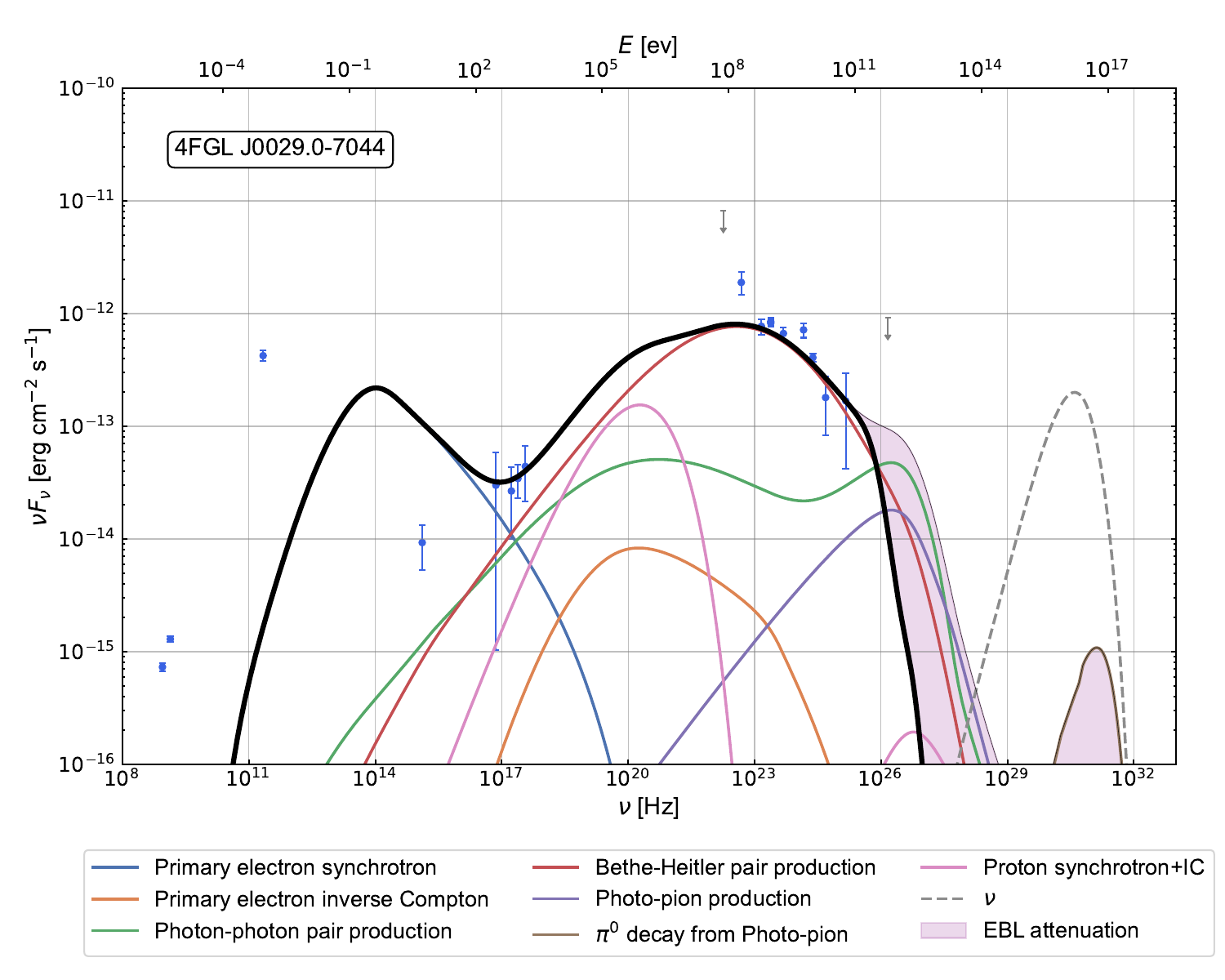}
\caption{
The SED modeling plots for the 103 NBCs.
The blue curve stands for the electron synchrotron emission;
the green curve stands for the photon-photon pair production;
the red curve stands for the Bethe-Heitler pair production;
the purple curve stands for the photon-pion production;
the brown curve stands for the proton synchrotron emission;
the pink curve stands for the $\pi^{0}$ decay;
the dashed-gray curve stands for the neutrino emission;
the pink shade stands for the attenuation from the extragalactic background light (EBL);
and the solid-black line represents the summed spectrum.
Only the first two plots (4FGL J0006.3-0620 and 4FGL J0029.0-7044) of the NBCs are shown here, the entire version will be displayed online.
}
\label{SED_plot}
\end{figure}

Our sample consists of 5 flat-spectrum radio quasars (FSRQs), 13 blazar candidates of uncertain type (BCUs) and 85 BL Lacertae objects (BL Lacs). 
The sources span a redshift range from $z = 0.01$ to $z = 2.340547$. 
The broadband SEDs of all sources are presented in Figure~\ref{SED_plot}, while the parameters describing the emission region, as well as the electron and proton energy distributions, are summarized in Table~\ref{SED_para}.

\begin{figure}[htbp]
\centering
\includegraphics[scale=0.6]{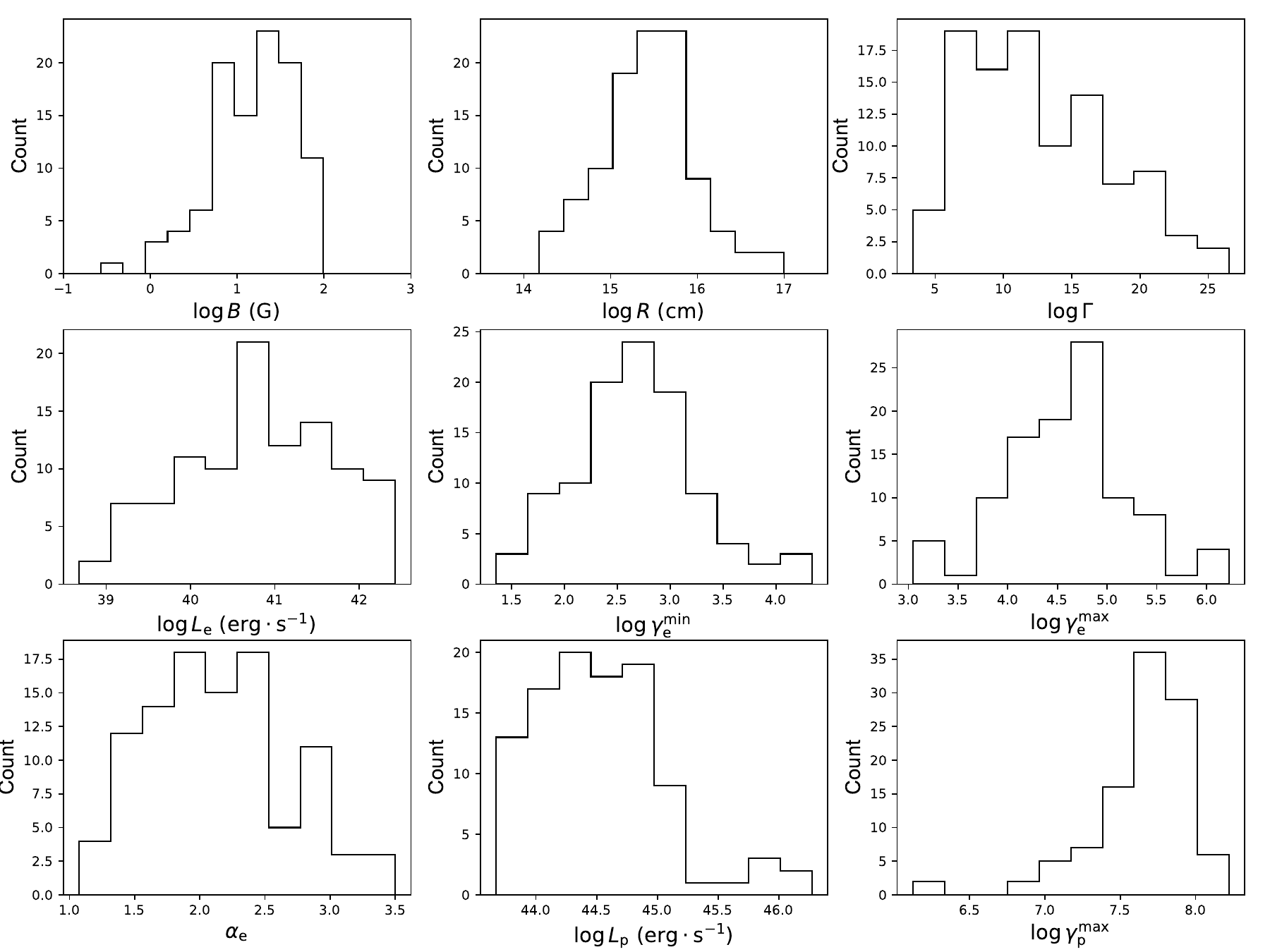}
\caption{The distribution of SED modeling parameters}
\label{Para_hist}
\end{figure}

The distributions of the nine free parameters in our model are shown in Figure~\ref{Para_hist}. 
We find that the magnetic field strength tends to approach the assumed upper limit, with 68 out of 103 sources (66.0\%) exhibiting $\log B \geq 1.0$. 
This behavior arises naturally from our modeling strategy, in which the low-energy hump of the blazar SED is attributed to synchrotron radiation from relativistic electrons, with the electron energy assumed to be primarily dissipated through the synchrotron process. 
Consequently, our model requires relatively strong magnetic field strengths.
To suppress the inverse Compton contribution and maximize the neutrino output, a relatively strong magnetic field is required to efficiently dissipate the electron energy through synchrotron emission, thereby allowing protons to dominate the $\gamma$-ray emission.
The average energy ratio between protons and electrons is $\log (P_{\rm p}/P_{\rm e}) \simeq 3.73 \pm 0.78$, and the one between protons and magnetic field is $\log (P_{\rm p}/P_{\rm B}) \simeq 1.67 \pm 1.03$.
Therefore, the power carried by the relativistic protons dominate over both the power carried by relativistic electron and magnetic field.

\begin{table}[htbp]
\centering
\caption{The SED modeling parameters}
\label{SED_para}
\begin{tabular}{lcccccccccccc}
\hline
4FGL Name &	Class & z & $\log R_{\rm em}$ &	$\log B$	& $\Gamma$ & $\log L_{\rm e}$ & $\log \gamma_{\rm e}^{\rm min}$ & $\log \gamma_{\rm e}^{\rm max}$	& $\alpha_{\rm e}$ &$\log L_{\rm p}$	& $\log \gamma_{\rm p}^{\rm max}$ & $\nu F_{\nu}$ \\
 & & & cm & G & & ${\rm erg/s}$ & & & & ${\rm erg/s}$ & & ${\rm erg/cm^{2}/s}$ \\    
(1)	& (2) &	(3)	& (4) &	(5)	& (6) &	(7)	& (8) &	(9)	& (10) & (11) &	(12) & (13)\\ 
\hline
J0006.3-0620	&	bll	&	0.346676	&	17.00	&	0.449	&	10.29	&	41.95	&	2.31	&	3.82	&	2.90	&	45.81	&	7.53	&	1.33E-14	\\
J0029.0-7044	&	bll	&	0.27	&	15.06	&	0.947	&	7.17	&	40.71	&	2.80	&	5.26	&	2.81	&	45.23	&	7.36	&	2.00E-13	\\
J0045.7+1217	&	bll	&	0.255	&	14.82	&	1.294	&	12.86	&	40.65	&	2.27	&	5.08	&	1.99	&	43.90	&	7.89	&	2.03E-12	\\
J0121.8-3916	&	bll	&	0.3	&	15.84	&	0.720	&	11.50	&	40.75	&	3.46	&	5.30	&	2.40	&	45.13	&	7.54	&	8.67E-13	\\
J0123.1+3421	&	bll	&	0.27	&	15.55	&	0.800	&	17.70	&	40.74	&	4.07	&	4.86	&	1.08	&	44.64	&	7.22	&	2.31E-12	\\
\hline
\end{tabular}
\tablecomments{
Column (1) 4FGL name;
column (2) classification;
column (3) redshift;
column (4) logarithmic dissipation region size, $R_{\rm em}$, in units of cm;
column (5) magnetic field strength, $B$, in units of G;
column (6) bulk Lorentz factor;
column (7) luminosity of injected relativistic electrons;
column (8) the minimum energy of electron population;
column (9) the maximum energy of electron population;
column (10) spectral index of the electron distribution;
column (11) luminosity of injected relativistic protons;
column (12) the maximum energy of proton population;
column (13) flux of neutrino emission;
Only five items are displayed, the entire table is available in machine-readable form.
}
\end{table}

\subsection{The correlation between neutrino emission and electromagnetic radiation}
\begin{figure}[htbp]
\centering
\includegraphics[scale=0.7]{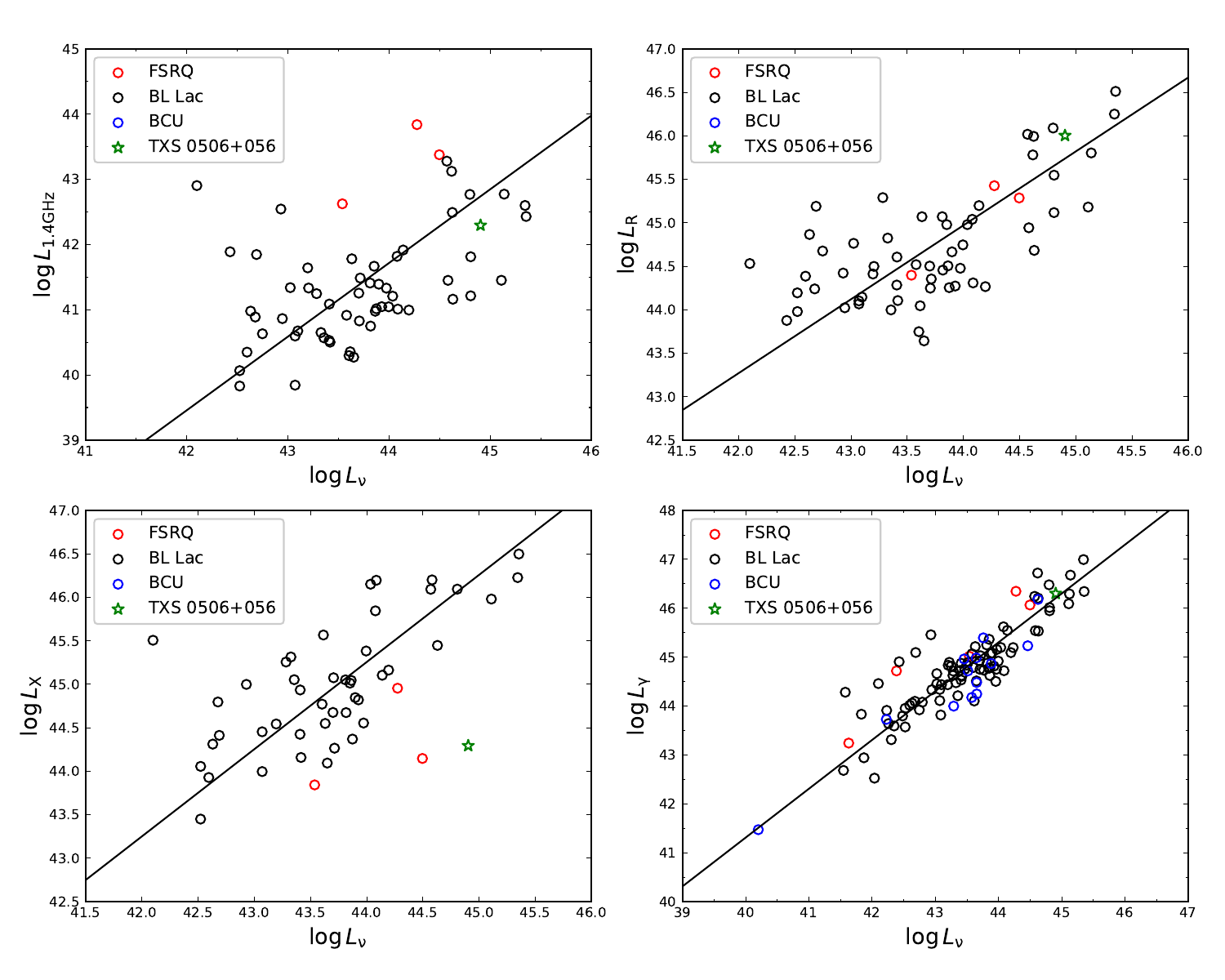}
\caption{Electromagnetic luminosity as a function of neutrino luminosity.
The red circles represent FSRQs, the black circles denote BL Lac objects, and the blue circles indicate BCUs. 
The green star marks TXS~0506+056, which is excluded from the linear regression, and the solid black line shows the best-fit linear regression.}
\label{P_comp}
\end{figure}

In this work, we collected the mass of central supermassive black hole and luminosity of accretion disk from \citet{Paliya2021}, radio flux density, optical magnitude, and X-ray flux from the BZCAT\footnote{https://www.ssdc.asi.it/bzcat/}, $\gamma$-ray photon flux, and corresponding photon index from 4FGL.
These data are all listed in Table \ref{multi}.
The multi-wavelength intensities are all converted into luminosity through 
\begin{equation}
    L = 4\pi d_{\rm L}^{2} (1+z)^{\alpha -1} F,
\end{equation}
in which $F$ is the flux in a giving band and $\alpha$ is the corresponding spectral index, where $d_{\rm L} = \frac{c}{H_{\rm 0}}\int^{1+z}_{1}\frac{1}{\sqrt{\Omega_{\rm m}x^{3}+1-\Omega_{\rm m}}}dx$ is a luminosity distance \citep{Komatsu2011}.
During the calculation the radio spectral index $\alpha_{\rm r}=0$ is applied for all sources in our sample; 
the optical spectral index $\alpha_{\rm o} =0.5$ is applied for BL Lacs and $\alpha_{\rm o} =1$ is applied for the rest;
the X-ray spectral index $\alpha_{\rm x} =1.3$ is applied for BL Lacs, $\alpha_{\rm x} =0.78$ is applied for FSRQs, and $\alpha_{\rm x} =1.05$ is applied for the BCUs;
and the $\gamma$-ray spectral index $\alpha_{\gamma} = \alpha_{\rm ph} -1$, these applications are also employed in literature \citep[e.g.,][]{Donato2001A&A375, Abdo2010ApJ716, Fan2016ApJS}.
The optical magnitude is corrected by the foreground galactic extinction with the value collected from NASA/IPAC Extragalactic Database (NED)\footnote{https://ned.ipac.caltech.edu/forms/byname.html}. 
We explored the correlation between neutrino luminosity ($\log L_{\nu}$) and 1.4 GHz radio luminosity ($\log L_{\rm 1.4GHz}$), optical $R$ band luminosity ($\log L_{\rm R}$), X-ray luminosity ($\log L_{\rm x}$) and $\gamma$-ray luminosity ($\log L_{\rm \gamma}$), fitted the correlation in the form of $y = \beta x + a$ with the ordinary least squares (OLS) bisector regression.
The results are illustrated in Figure \ref{P_comp} and in Table \ref{corr}, considering that the luminosity is correlated with the redshift, we removed its influence to achieve the pure correlation of the luminosities through a partial correlation analysis:
\begin{equation}
    r_{ij,k} \frac{r_{ij} - r_{ik}r_{jk}}{\sqrt{(1-r_{ik}^{2})(1-r_{jk}^{2})}},
\end{equation}
in which $r_{ij}$, $r_{ik}$ and $r_{jk}$ are the correlation coefficients between the pair of variables $i,j \, {\rm and}\,k$, the variables $i$ and $j$ represent luminosities, and the variable $k$ represents redshift.

\begin{table}[htbp]
\centering
\caption{The central engine and multi-wavelength emission}
\label{multi}
\begin{tabular}{lccccccc}
\hline
4FGL Name & $\log M_{\odot}$ & $\log L_{\rm d}$ & $f_{\rm 1.4 \, GHz}$ & $m_{\rm R}$ & $F_{\rm x}$ & Flux1000 & $\alpha_{\rm ph}$ \\
 &  & ${\rm erg/s}$ & mJy & mag & ${\rm erg/cm^{2}/s}$ & ${\rm ph/cm^{2}/s}$ & \\
(1)	         &	(2)	    &(3)   & (4)   & (5)  &	(6)  &	(7)	&	(8)	    	\\ 
\hline
J0006.3-0620	&	8.93	&	44.52	&	2051	&	17.9	&	0.82	&	1.21E-10	&	2.17	\\
J0029.0-7044	&		&		&	94	&	16.7	&		&	3.93E-10	&	2.31	\\
J0045.7+1217	&	8.82	&	44.18	&	104	&	17.4	&	0.56	&	9.85E-10	&	1.95	\\
J0121.8-3916	&	8.58	&	44.39	&		&		&		&	2.50E-10	&	1.93	\\
J0123.1+3421	&		&		&	44	&	17.9	&	25.3	&	2.30E-10	&	1.69	\\
\hline
\end{tabular}
\tablecomments{
Column (1) 4FGL name;
column (2) black hole mass in unit of solar mass;
column (3) accretion disk luminosity;
column (4) flux density at 1.4 GHz;
column (5) optical R band magnitude;
column (6) X-ray flux in the range of 0.1 - 2.4 keV;
column (7) the $\gamma$-ray integral photon flux from 1 to 100 GeV; 
column (8) the $\gamma$-ray photon index of `Flux1000';
Only five items are displayed, the entire table is available in machine-readable form.
}
\end{table}

\begin{table}[htbp]
\centering
\caption{Regression results between neutrino luminosity and the electromagnetic luminosity.}
\label{corr}
\begin{tabular}{lccccc}
\hline
Correlation & N & $\beta$ & $\alpha$ & r & p  \\
(1) & (2) & (3) & (4) & (5) & (6)  \\
\hline
$\log L_{\rm 1.4\,GHz}  \, {\rm vs} \, \log L_{\nu}$  & 60  & $1.13 \pm 0.10$  & $-8.00 \pm 4.22$ & -0.13  & 0.33  \\
$\log L_{\rm R}  \, {\rm vs}      \, \log L_{\nu}$  & 60  & $0.85 \pm 0.07$  & $7.56 \pm 2.85$ &  0.30  & 0.02  \\
$\log L_{\rm x}  \, {\rm vs}      \, \log L_{\nu}$  & 48  & $1.00 \pm 0.08$  & $1.12 \pm 3.50$  &  0.52  & $1.8 \times 10^{-4}$  \\ 
$\log L_{\gamma} \, {\rm vs} \,      \log L_{\nu}$  & 103 & $1.00 \pm 0.04$  & $1.37 \pm 1.89$  &  0.78  & $9.1 \times 10^{-22}$ \\
\hline
\end{tabular}
\tablecomments{
column (1) gives the name of correlation;
column (2) is the number of the sources;
column (3) gives the slope;
column (4) gives the intercept;
column (5) gives the partial Pearson correlation coefficient;
column (6) gives the chance probability.
}
\end{table}

The partial Pearson correlation analysis indicates positive correlations between $\log L_{\nu}$ and $\log L_{\rm R}$, $\log L_{\rm X}$, and $\log L_{\gamma}$. 
We find that the strength of the correlation increases with the photon energy of the considered electromagnetic bands. 
Among the four bands examined, the correlation between $\log L_{\nu}$ and $\log L_{\gamma}$ is the tightest, followed by that between $\log L_{\nu}$ and $\log L_{\rm X}$.
Within the framework of the hadronic model, the high-energy hump of the blazar SED is primarily produced by proton synchrotron radiation and Bethe–Heitler pair production. 
Proton synchrotron emission mainly contributes to the correlation regarding X-ray band, while Bethe–Heitler processes dominate the $\gamma$-ray emission.
Both mechanisms involve interactions of high-energy protons, and the resulting high-energy photons can act as seed photons for photo-pion production, ultimately leading to neutrino emission.
There is weak/moderate correlation between and $\log L_{\nu}$ and $\log L_{\rm R}$, and no correlation for the $\log L_{\nu}$ and $\log L_{\rm 1.4 \, GHz}$.

\subsection{The correlation between neutrino emission and the central engine}
\begin{figure}[htbp]
\centering
\includegraphics[scale=0.7]{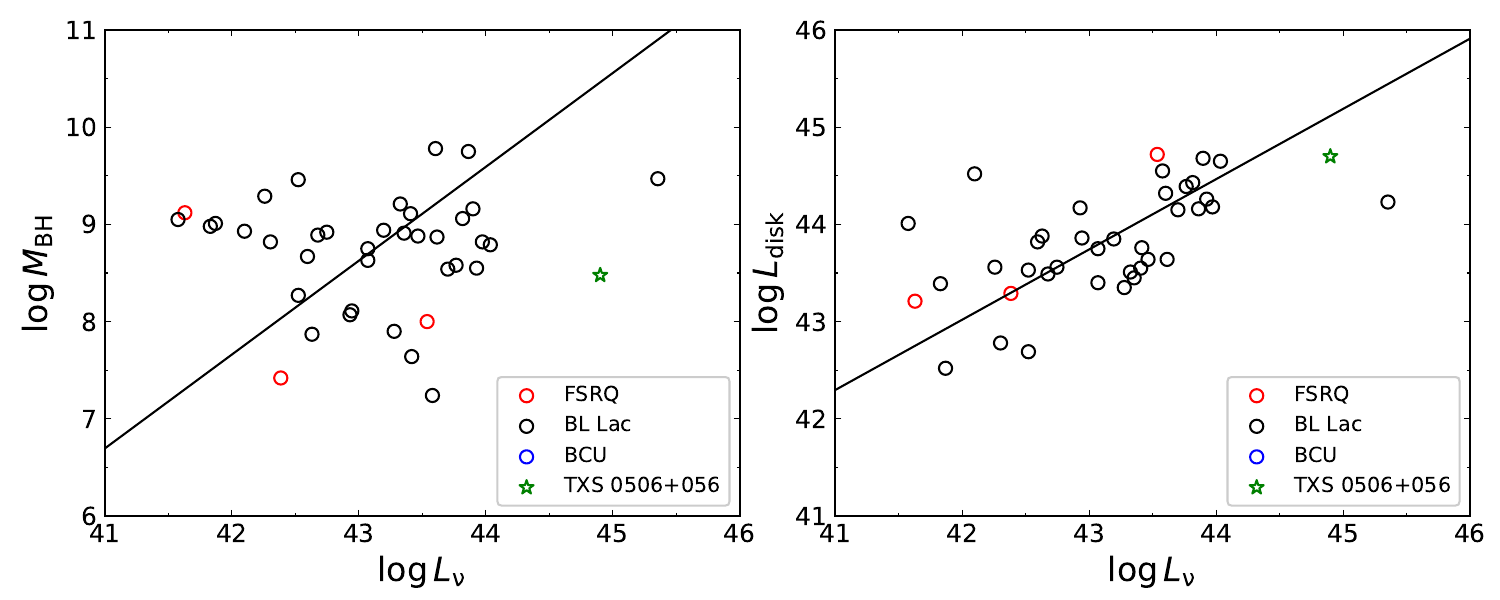}
\caption{Central engine as a function of neutrino luminosity. 
The symbols used for the data points and the regression lines are the same as those adopted in Figure \ref{P_comp}.}
\label{NMLd}
\end{figure}

We investigate the potential connection between neutrino emission and the central engine properties, specifically the mass of the central supermassive black hole ($M_{\rm BH}$) and the accretion disk luminosity ($L_{\rm disk}$). 
These quantities are adopted from \citet{Paliya2021}, where $M_{\rm BH}$ and $L_{\rm disk}$ are estimated primarily using emission-line properties, while absorption-line features and bulge luminosity are additionally employed to constrain $L_{\rm disk}$.
The correlations are examined using linear relations of the form $y = \beta x + \alpha$, derived through ordinary least-squares (OLS) bisector regression. 
The regression results are presented in Figure~\ref{NMLd} and summarized in Table~\ref{corr_NMLd}. 
In addition, a partial correlation analysis is performed to evaluate the intrinsic correlation between $\log L_{\nu}$ and $\log L_{\rm disk}$.

No significant linear correlation is found for either $\log M_{\rm BH} \, {\rm vs} \, \log L_{\nu}$ or $\log L_{\rm disk} \, {\rm vs} \, \log L_{\nu}$. 
For $\log M_{\rm BH}$ and $\log L_{\nu}$, the Pearson correlation coefficient is $r = 0.06$ with a chance probability of $p = 0.72$, indicating the absence of a meaningful correlation.
The relationship between $\log L_{\rm disk}$ and $\log L_{\nu}$ appears to be more complex. 
After accounting for the redshift dependence inherent in luminosity–luminosity correlations, the partial correlation analysis yields $r = 0.24$ and $p = 0.16$, suggesting that the correlation is not statistically significant. 
Nevertheless, a weak positive trend between $\log L_{\rm disk}$ and $\log L_{\nu}$ is visually apparent in the right panel of Figure~\ref{NMLd}, which may hint at an underlying connection that cannot be firmly established with the current sample.

\begin{table}[htbp]
\centering
\caption{Regression results of the central engine and neutrino luminosity.}
\label{corr_NMLd}
\begin{tabular}{lccccc}
\hline
Correlation & N & $\beta$ & $\alpha$ & r & p  \\
(1) & (2) & (3) & (4) & (5) & (6)  \\
\hline
$\log M_{\rm BH}   \, {\rm vs} \, \log L_{\nu}$ & 38  & $0.97 \pm 0.09$ & $-32.89 \pm 3.84$ & 0.06  & 0.72  \\
$\log L_{\rm disk} \, {\rm vs} \, \log L_{\nu}$ & 38  & $0.72 \pm 0.09$ & $12.65 \pm 3.91$ & 0.24  & 0.16  \\
\hline
\end{tabular}
\tablecomments{
column (1) gives the name of correlation;
column (2) is the number of the sources;
column (3) gives the slope;
column (4) gives the intercept;
column (5) gives the Pearson correlation coefficient;
column (6) gives the chance probability;
}
\end{table}

\section{Discussion}
\subsection{The hadronic model of this work}
In hadronic models, both electrons and protons are accelerated to relativistic energies, with protons exceeding the threshold for $p\gamma$ photo-pion production on the ambient soft photon field.
The lower-energy hump is still dominated by electron synchrotron radiation, whereas the higher-energy hump is mainly produced by proton synchrotron emission, $\pi^{0}$ decay, and electromagnetic cascades initiated by secondary particles \citep{Mannheim1992A&A253, Aharonian2000NewA5, Mucke2003APh, Bottcher2013}.
In this work, we employ a hadronic model to reproduce the high-energy component of our NBCs, using multi-wavelength data collected from public databases.
Compared to previous hadronic models in the literature \citep[e.g.,][]{Zech2017A&A602, Rodrigues2024A&A681}, in which the GeV emission observed by Fermi-LAT is predominantly attributed to proton synchrotron radiation, while cascade emission accounts for the radiation beyond the GeV band, our modeling suggests a different scenario.
In our case, the proton synchrotron component is shifted to the MeV range for most NBCs, whereas the GeV emission is primarily produced by synchrotron radiation from Bethe–Heitler pairs.
The role of Bethe–Heitler pair production in shaping the $\gamma$-ray peak has been investigated in detail by \citet{Karavola2024JCAP07}, who showed that blazars with the lower-energy hump located at $10^{14\text{–}15.5},\mathrm{Hz}$ and the higher-energy hump peaking at $10^{22\text{–}25},\mathrm{Hz}$ are more likely to have their high-energy emission dominated by Bethe–Heitler processes. 
Our sample broadly falls within this regime, supporting such an interpretation.
Compared to their results, our modeling is characterized by, on average, a larger emission region size ($R$) and a smaller Doppler factor. 
In addition, we adopt a fixed proton spectral index of 1, which further distinguishes our parameter setup from previous studies.
With such parameter setup, we have a proton synchrotron bump mostly in the MeV range with a larger intensity than the Bethe-Heitler pair bump in the GeV range, which will be discussed in Section 4.3.

Hadronic models offer several advantages in addressing challenges encountered in one-zone leptonic models, such as extremely rapid variability and the requirement of very high bulk Lorentz factors \citep{Aharonian2007ApJ, Begelman2008MNRAS, Ouyang2025ApJ980}, very-high-energy (VHE) excess emission \citep{Albert2008, Bottcher2013, Cheng2022ApJ925}, and neutrino production in blazars \citep{Cerruti2019MNRAS483, Xue2021ApJ906}.
Nevertheless, hadronic models are often challenged by significant deviations from equipartition between the dominant particle component and the magnetic field \citep[e.g.,][]{Bottcher2013}, as well as by the requirement of super-Eddington jet powers \citep[e.g.,][]{Cheng2022ApJ925}.
In this work, we find that the power of relativistic particles dominate over the magnetic field by 1–2 orders of magnitude.
We define the equipartition parameter as $\epsilon_{pB} \equiv P_{\rm p}/P_{\rm B}$, an average value of $\log \epsilon_{pB} \simeq 1.67 \pm 1.03$ is found for the NBCs in our sample.
The maximum value, $\log \epsilon_{pB} = 5.27$, is found for 4FGL J0509.9-6417, corresponding to a relatively weak magnetic field strength of $\log B = -0.57$.
The minimum value, $\log \epsilon_{pB} = -0.36$, is obtained for 4FGL J1719.3+1205, where comparable powers are inferred for the magnetic field ($\log P_{\rm B} = 46.35$) and protons ($\log P_{\rm p} = 45.89$).
Among the 103 NBCs in our sample, 10 sources exhibit $-0.5 \leq \log \epsilon_{pB} \leq 0.5$, indicating approximate energy equipartition between particles and the magnetic field.
For the 39 sources with available black hole mass estimates, 14 sources show total jet powers exceeding the Eddington luminosity, $L_{\rm Edd} = 1.26 \times 10^{38} (M_{\rm BH}/M_{\odot})~{\rm erg~s^{-1}}$.
The ratio $P_{\rm tot}/L_{\rm Edd}$ ranges from a minimum of 0.01 to a maximum of 2.55, suggesting that a significant fraction of the sources operate at or above the Eddington limit.
The problem of exceed the Eddington limit can be reconciled by involving multi-zone models.
Recently, for instance, \citet{Rodrigues2026A&A706} proposed a physically motivated, continuous multi-zone blazar jet lepto-hadronic model, where particle acceleration and radiation evolve along the jet, and successfully calculated SEDs of TXS 0506+056 and PKS 0605-085.

\subsection{The correlations between neutrino emission and other parameters}
Among the examined correlations between neutrino luminosity and electromagnetic luminosity, the relationship between $\log L_{\nu}$ and $\log L_{\rm 1.4,GHz}$ yields a chance probability of $p = 0.33$, indicating no statistically significant correlation between neutrino emission and radio-band emission.
In contrast, the correlations of $\log L_{\nu}$ with $\log L_{\rm X}$ and $\log L_{\nu}$ with $\log L_{\gamma}$ are both strong. 
This behavior is expected if the X-ray and $\gamma$-ray emissions originate from hadronic processes.
Considering that the X-ray and $\gamma$-ray fluxes are independently compiled from BZCAT and represent the average emission levels in these bands, thus the observed significant correlations imply that a hadronic origin of the X-ray and $\gamma$-ray emission in blazars cannot be completely excluded.

An intriguing result among these correlations is the weak/moderate correlation between $\log L_{\nu}$ and $\log L_{\rm R}$.
Within our SED modeling framework, the optical emission is primarily attributed to synchrotron radiation from relativistic electrons, a purely electromagnetic process that does not produce neutrinos.
Nevertheless, several factors may contribute to the observed correlation.
First, this trend may arise from sample selection effects and could be modified or weakened when a larger and more complete sample is considered.
If the correlation has a physical origin, it may arise from multiple contributing mechanisms.
In particular, the optical emission in the $R$ band is composed of radiation from both the jet and the accretion disk, with the jet component generally dominating.

\begin{figure}[htbp]
\centering
\includegraphics[scale=0.7]{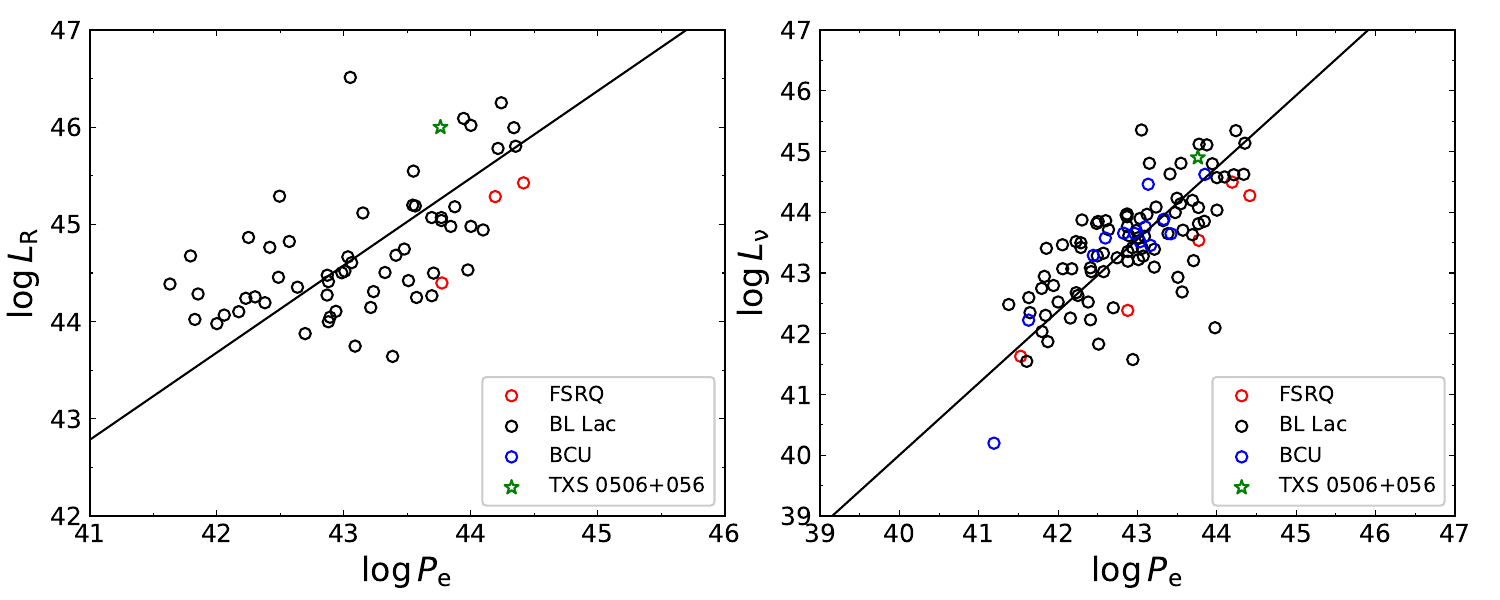}
\caption{Optical luminosity (left) and neutrino luminosity (right) as a function of the electron luminosity.
The symbols used for the data points and the regression lines are the same as those adopted in Figure \ref{P_comp}.}
\label{NPR}
\end{figure}

A jet-related origin of this correlation is a natural consequence if the optical emission provides seed photons for photo-pion production, thereby leading to subsequent neutrino emission, provided that protons are accelerated to sufficiently high energies.
The correlations between the jet electron power ($\log P_{\rm e}$) and $\log L_{\rm R}$, as well as between $\log P_{\rm e}$ and $\log L_{\nu}$, are shown in Figure~\ref{NPR}, with the corresponding regression results listed in Table~\ref{corr_NPR}.
These correlations support our proposed interpretation.
In the left panel of Figure~\ref{NPR}, we find a positive correlation between $\log L_{\rm R}$ and $\log P_{\rm e}$, where $P_{\rm e} \simeq \Gamma^{2} L_{\rm e}$, indicating that the optical emission is indeed dominated by synchrotron radiation from relativistic electrons in the jet.
In the right panel of Figure~\ref{NPR}, a positive correlation is also observed between $\log L_{\nu}$ and $\log P_{\rm e}$, suggesting that synchrotron photons produced by relativistic electrons act as seed photons for photo-pion ($p\gamma$) interactions.

\begin{table}[htbp]
\centering
\caption{The linear regression results of electron power and the neutrino luminosity, and optical luminosity.}
\label{corr_NPR}
\begin{tabular}{lccccc}
\hline
Correlation & N & $\beta$ & $\alpha$ & r & p  \\
(1) & (2) & (3) & (4) & (5) & (6)  \\
\hline
$\log L_{\rm R}  \, {\rm vs} \, \log P_{\rm e}$ & 60  & $0.90 \pm 0.09$ & $6.00 \pm 3.73$  & 0.58  & $1.3 \times 10^{-6}$   \\
$\log L_{\nu}    \, {\rm vs} \, \log P_{\rm e}$ & 103 & $1.18 \pm 0.08$ & $-7.37 \pm 3.28$ & 0.72  & $1.1 \times 10^{-17}$  \\
\hline
\end{tabular}
\tablecomments{
column (1) gives the name of correlation;
column (2) is the number of the sources;
column (3) gives the slope;
column (4) gives the intercept;
column (5) gives the Pearson correlation coefficient;
column (6) gives the chance probability;
}
\end{table}

In addition, a disk-related origin for this correlation cannot be excluded, neutrino emission may be related to optical emission through the accretion disk if a fraction of the optical $R$-band emission in blazars originates from the disk.
In this case, the optical $R$-band luminosity can serve as a proxy for the disk luminosity, such that stronger $R$-band emission corresponds to a more powerful accretion disk.
Under this assumption, the observed correlation between $\log L_{\rm R}$ and $\log L_{\nu}$ can be interpreted as a manifestation of an underlying correlation between the disk luminosity, $\log L_{\rm disk}$, and the neutrino luminosity, $\log L_{\nu}$.
The correlation between $\log L_{\rm disk}$ and $\log L_{\nu}$, shown in Figure~\ref{NMLd} and quantified in Table~\ref{corr_NMLd}, suggests the presence of a weak positive trend.
If the relatively large $p$-value is primarily due to the limited sample size (38 sources), and if a positive correlation between these two quantities indeed exists, it may be understood in two possible ways.
On the one hand, photons associated with the accretion disk and/or corona —such as X-ray or optical photons— may serve as seed photons for photo-pion interactions \citep{Xue2021ApJ906, Murase2022ApJL941}.
On the other hand, blazar neutrino emission may originate from the inner regions of the accretion disk or the corona (i.e., the AGN core), provided that hadrons are accelerated together with leptons \citep{Begelman1990ApJ362, Inoue2019ApJ880}.
Previous studies have suggested that the AGN core itself could act as a site of high-energy neutrino production \citep{Stecker2013PRD88, Kalashev2015JETP120}, although such emission may be insufficient to account for the IceCube detection associated with TXS~0506+056 \citep{Fiorillo2025ApJ986}.

\subsection{The particle composition of blazar jet}
Leptonic models have long been established as the standard framework for explaining blazar emission, in which low-energy radiation is produced by electron synchrotron emission and high-energy photons arise from the IC scattering.
Motivated by the proton-dominated composition of cosmic rays, hadronic models were subsequently proposed, in which relativistic protons in jets interact with ambient photon fields to produce high-energy $\gamma$ rays and neutrinos.
Although hadronic models can naturally account for TeV emission and neutrino production, they often require jet powers exceeding the Eddington limit.
The detection of high-energy neutrinos from blazars has provided compelling multi-messenger evidence for the presence of relativistic hadrons in blazar jets.

\subsubsection{The emission in the MeV range}
In this work, we adopt a hadronic model to construct the broadband SEDs of 103 neutrino blazar candidates (NBCs) and to investigate the physical properties of blazar neutrino emission.
To obtain the highest possible neutrino flux, we impose the condition that the high-energy emission produced by the leptonic IC scattering is suppressed to a level three orders of magnitude lower than that arising from $p\gamma$ interactions.
To achieve this, the electron power is reduced while the magnetic field strength is increased, thereby enhancing synchrotron cooling of electrons.
Consequently, the proton energy budget must be increased to reproduce the observed GeV–TeV $\gamma$-ray emission.
As a result, synchrotron radiation from relativistic protons produces a distinct spectral bump in the broadband SED, located in the MeV energy range.
Indeed, for 99 out of the 103 NBCs in our sample, the proton synchrotron peak is found within the 0.1–100 MeV band with an average proton energy $\log\gamma_{\rm p}=7.89\pm0.34$.
Among these sources, 90 exhibit show significant proton synchrotron peak, with MeV-band proton synchrotron fluxes higher than the Bethe–Heitler pair production flux in the GeV band.
This distinctive spectral feature and the associated strong MeV emission are expected to be detectable by next-generation MeV $\gamma$-ray missions, such as e-ASTROGAM \citep{DeAngelis2018JHEAp19} and AMEGO \citep{McEnery2019BAAS51g}.
Moreover, the MeV bump can be tightly constrained through joint observations with future MeV $\gamma$-ray facilities and next-generation X-ray instruments, e.g., the enhanced X-ray Timing and Polarization telescope \citep[eXTP,][]{Zhou2025SCPMA6819507Z}, which provide broad energy coverage from the soft X-ray band (keV) through hard X-rays and extending into the MeV regime (up to $\sim$600 keV).
Such multi-wavelength observations will enable stringent tests of the hadronic scenario adopted in this work and provide valuable insights into the particle composition of blazar jets.

\subsubsection{The neutrino emission}
\begin{figure}[htbp]
\centering
\includegraphics[scale=0.6]{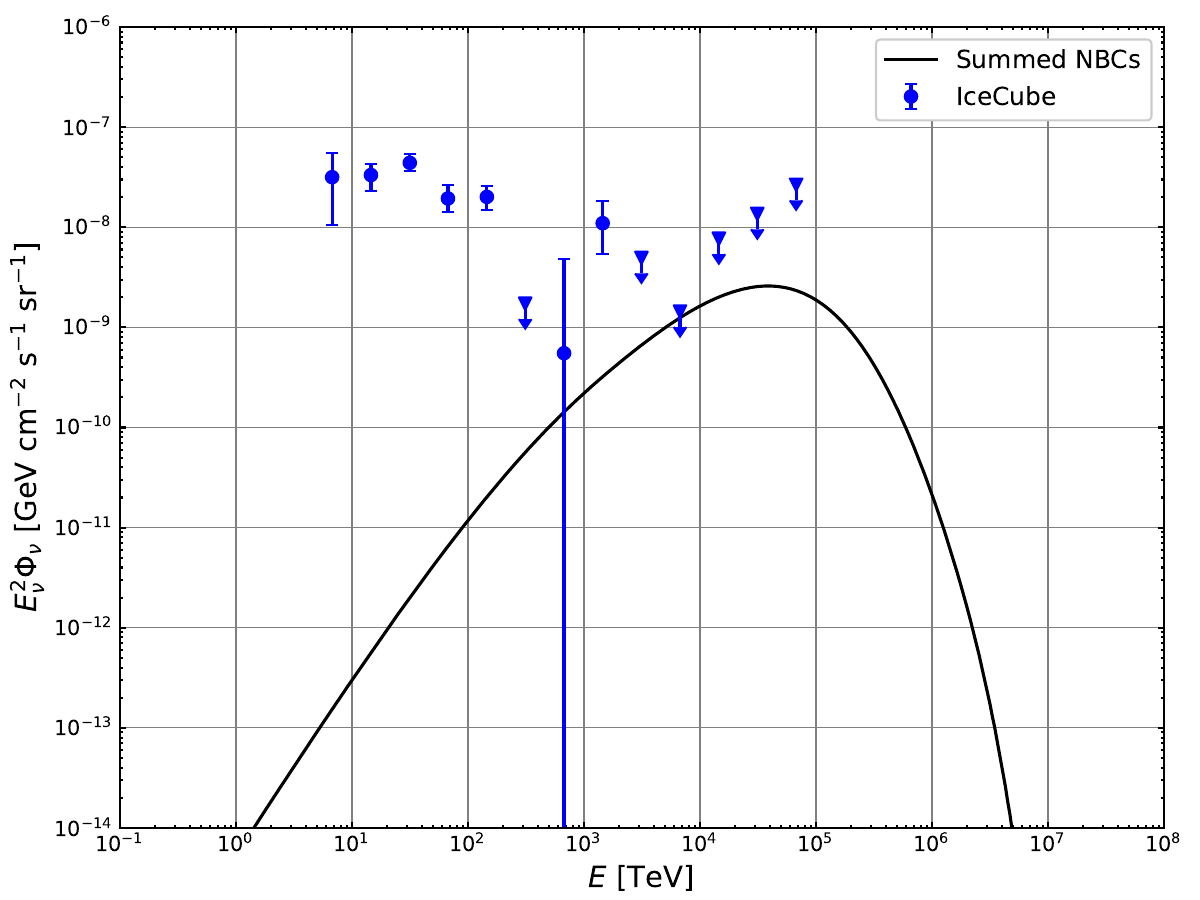}
\caption{The summed neutrino flux as a function of energy.
The blue data points and upper limits are represent the differential flux model best fit results for the 2010–2015 (six years) IceCube cascade data; the curve represents the summed neutrino flux from NBCs.}
\label{icelim}
\end{figure}

Defining the neutrino-to-$\gamma$-ray intensity ratio as $Y_{\nu\gamma} = L_{\nu}/L_{\gamma}$ \citep{Padovani2015MNRAS452}, where $L_{\nu}$ is the peak neutrino luminosity and $L_{\gamma}$ is the integrated $\gamma$-ray luminosity in the 1–100 GeV band. The results yield an average value of $\langle Y_{\nu\gamma} \rangle = 0.074 \pm 0.067$, with a maximum value of 0.32.
We find that 14 sources exhibit $Y_{\nu\gamma} > 0.13$, a value previously suggested as an approximate upper limit for blazars based on 7 yr IceCube ultra-high-energy (UHE) data \citep{Aartsen2016PRL117}. 
These sources tend to show relatively higher $L_{\nu}$ and lower $L_{\gamma}$ compared to the overall NBC sample.
In addition, we perform a rough estimate of the contribution of NBCs to the diffuse neutrino background using
\begin{equation}
    E_{\nu}^{2}\Phi_{\nu} = \frac{1}{4\pi}\frac{N_{\rm sample}}{N_{\rm all}}\, \Sigma_{i}(\nu_{i}F_{i}),
\end{equation}    
where $\nu_{i}F_{i}$ represents the neutrino flux of individual sources, $N_{\rm sample} = 103$ is the sample size in this work, and $N_{\rm all} = 199$ corresponds to the full NBC sample in Paper~I.
A comparison between our estimate and the IceCube stacking limit for blazars is shown in Fig~\ref{icelim}. 
The result indicates that our scenario, which explores the maximum neutrino emission capability of blazars, is not excluded by current observational constraints.

\begin{figure}[htbp]
\centering
\includegraphics[scale=0.6]{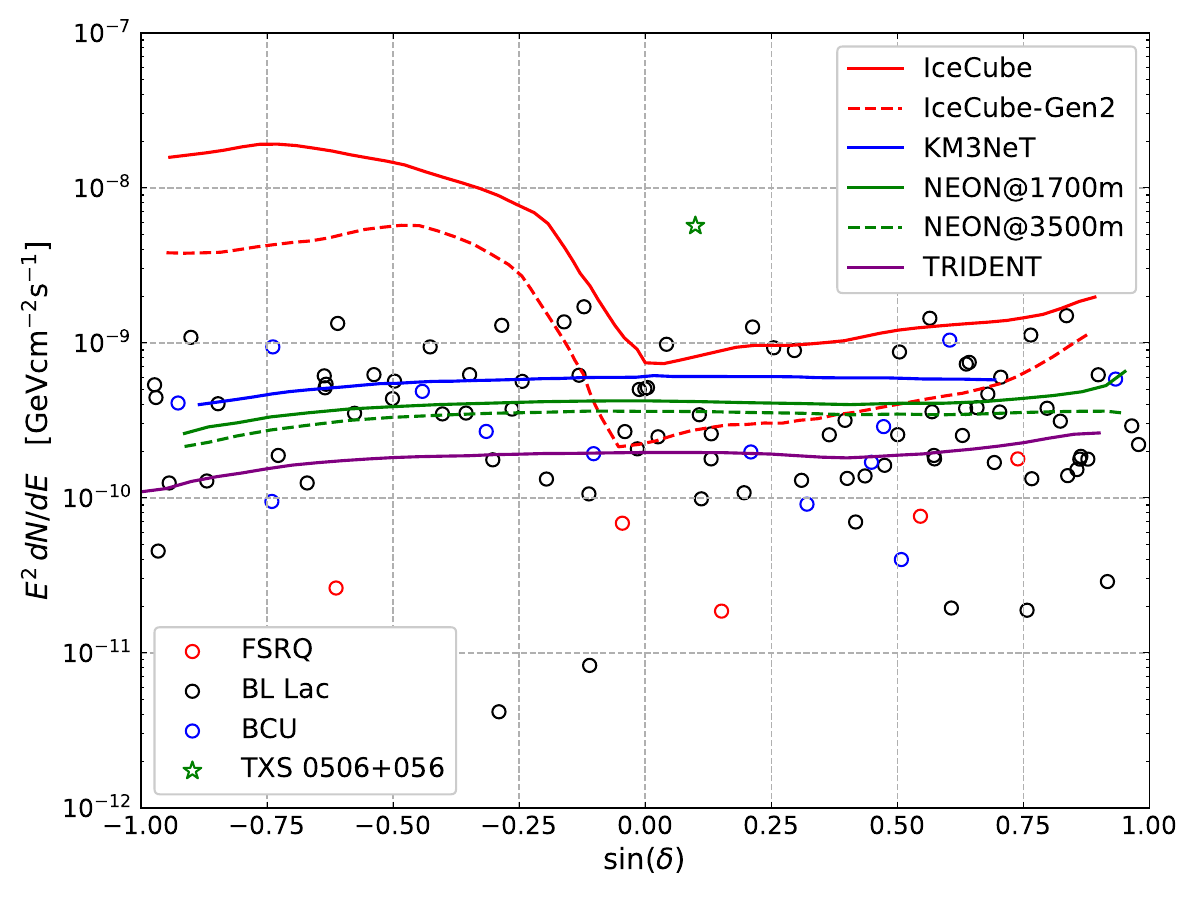}
\caption{The $5\sigma$ discovery potential of our neutrino blazar candidates (NBCs) for current and next-generation neutrino detectors as a function of declination, based on a 10-yr exposure.
The $5\sigma$ discovery potential curve in solid red for IceCube \citep{Aartsen2020PRL124};
the dashed red curve for IceCube-Gen2 \citep{Omeliukh2022icrc1184O};
the solid blue curve for KM3NeT \citep{Aiello2019APh111};
the solid and dashed green curves for NEON \citep{Zhang2025APh17103123};
the solid purple curve for TRIDENT \citep{TRIDENT2023};
the symbols used for the data points are the same as those adopted in the previous figures.}
\label{neu}
\end{figure}

Based on our SED results, the peak neutrino flux of these NBCs ranges from 1.55 PeV to 275.42 PeV, the range can be covered by the current and next generation of neutrino facilities, such as 
IceCube \citep{Aartsen2017JInst12}, 
the Cubic Kilometre Neutrino Telescope (KM3NeT) \citep{Adrian2016JPhG43}, 
the NEutrino Observatory in the Nanhai (NEON) \citep{Zhang2025APh17103123}, 
and IceCube Generation 2 (IceCube-Gen2) \citep{Aartsen2021}.
We compile the $5\sigma$ discovery potential fluxes of these neutrino facilities, which vary with declination, compare them with the peak neutrino fluxes predicted by our model in Figure~\ref{neu} and list the results in Table~\ref{neu_det}.
We find that three NBCs are potentially detectable by IceCube, 20 by IceCube-Gen2, 22 by KM3NeT, 35 by NEON@1700m, 45 by NEON@3500m, and 62 by TRIDENT.
Consequently, some of these sources could be confirmed as neutrino emitters by KM3NeT, NEON and TRIDENT in the coming few years.

\begin{figure}
\centering
\includegraphics[width=250pt]{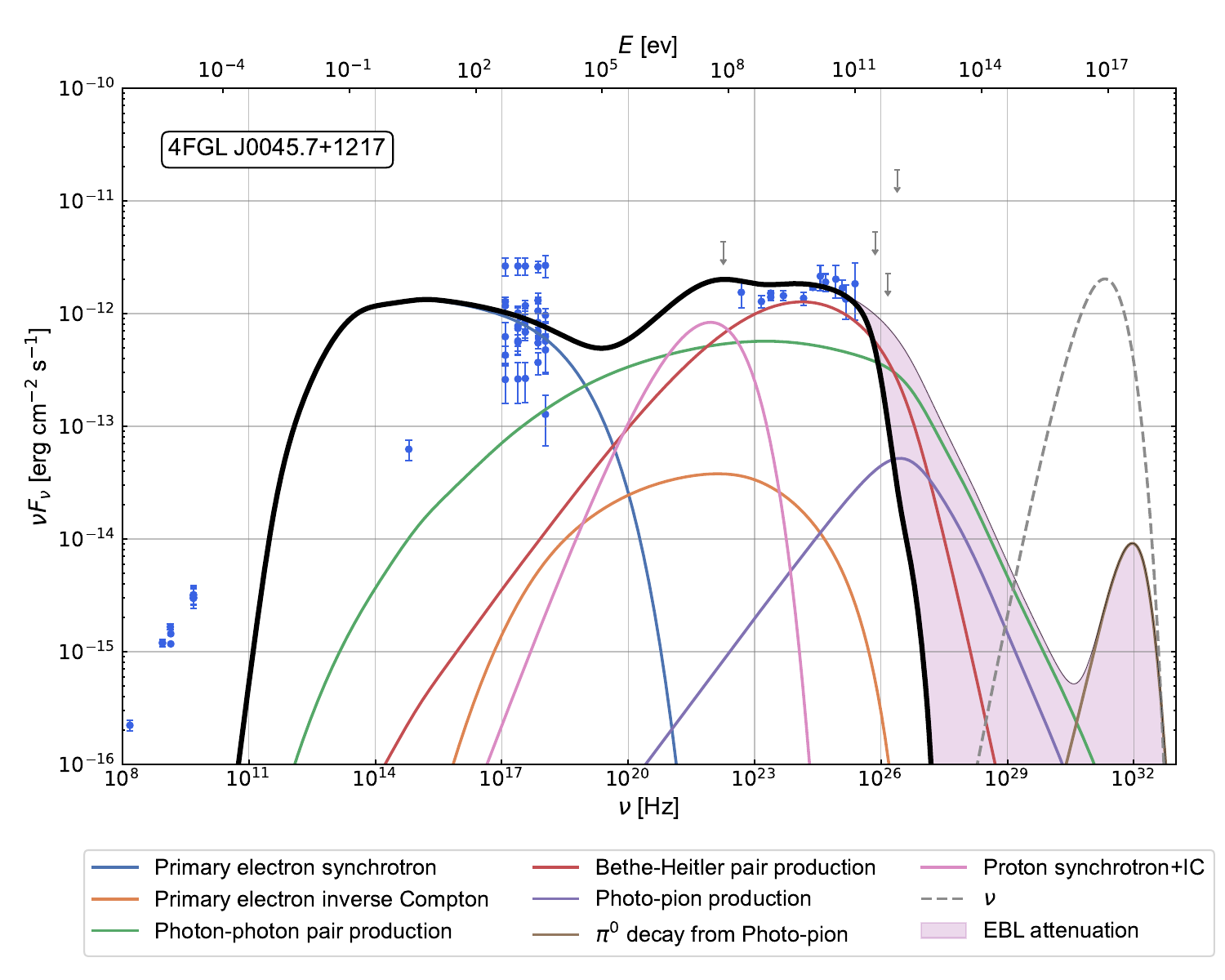}
\includegraphics[width=250pt]{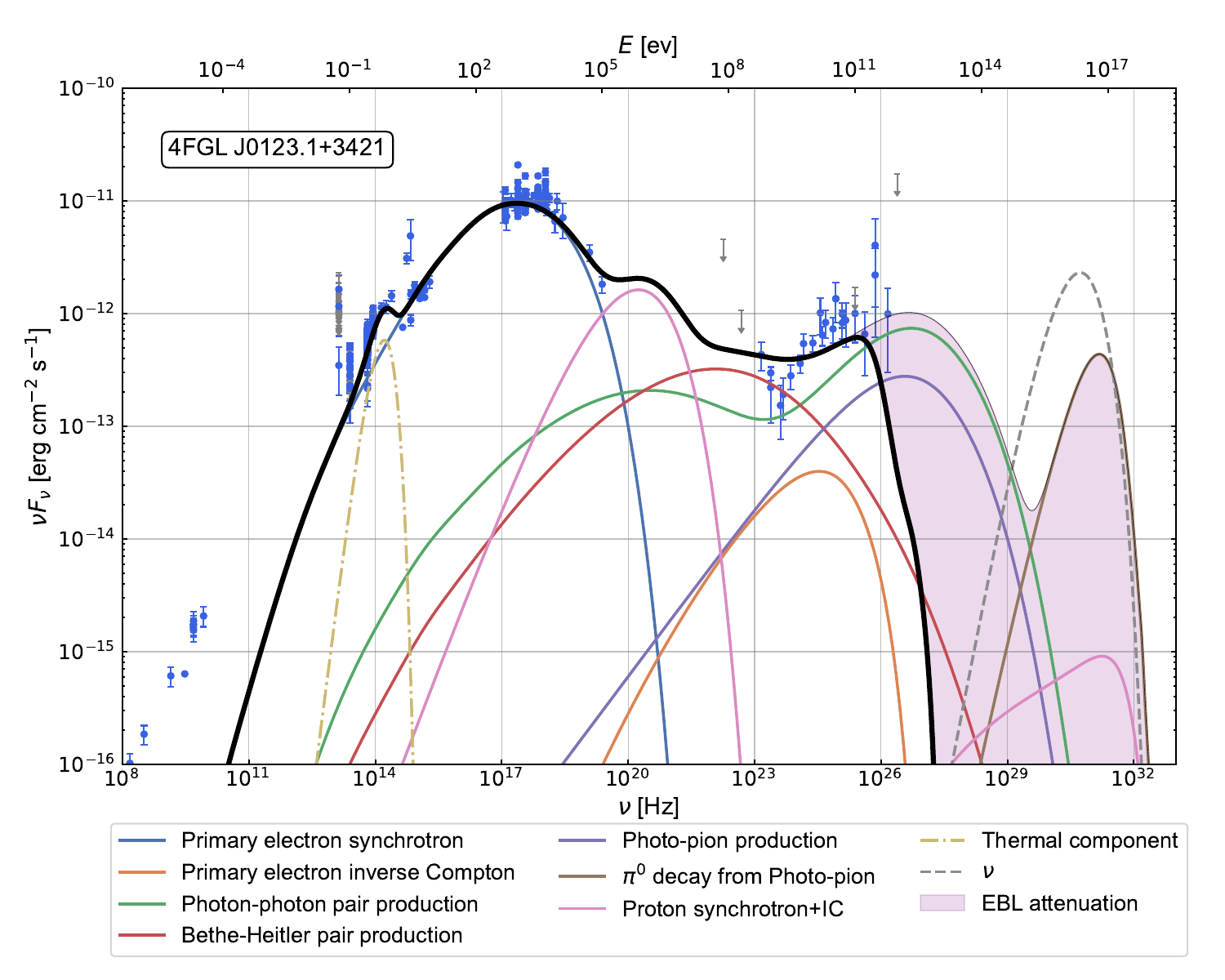}
\includegraphics[width=250pt]{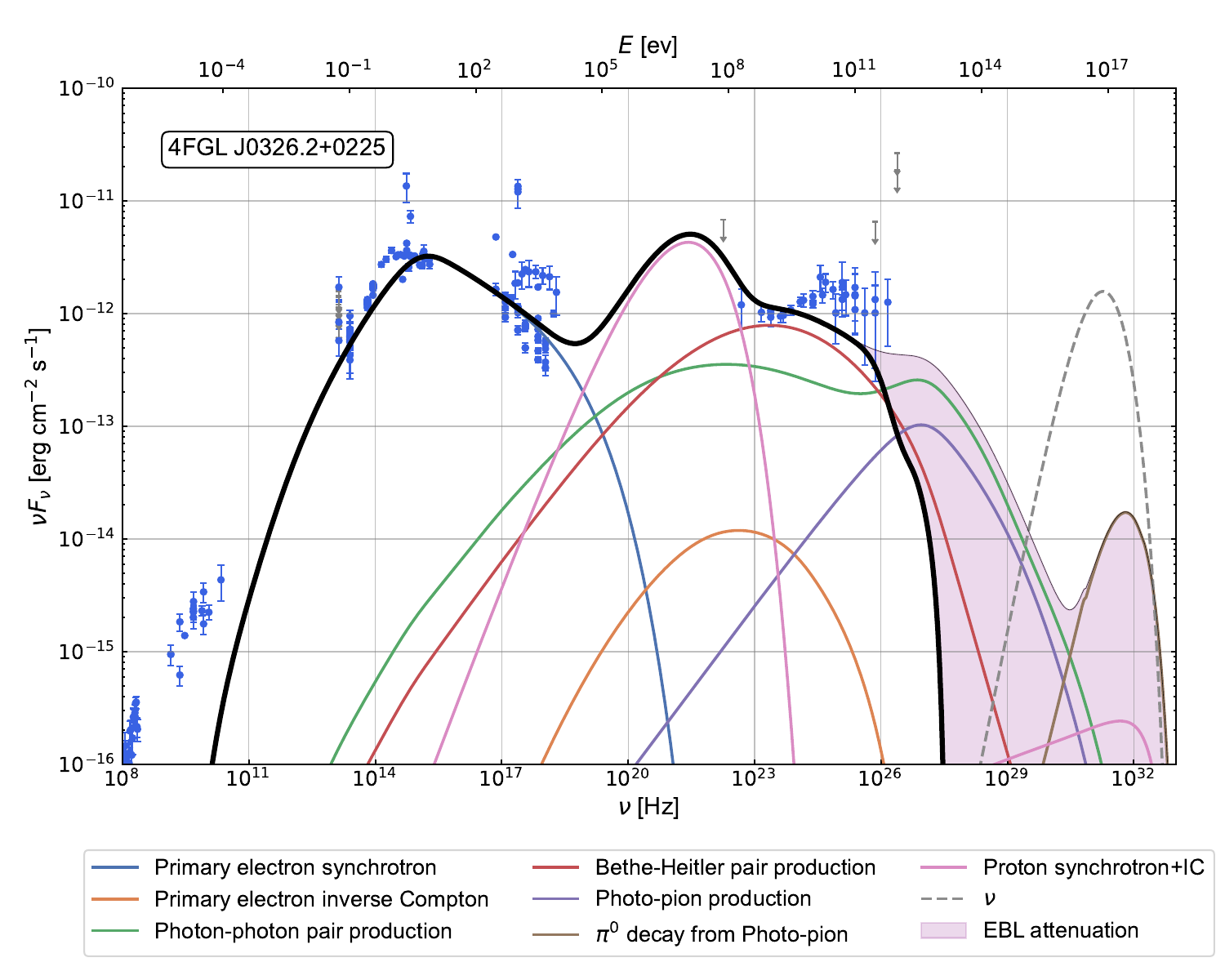}
\includegraphics[width=250pt]{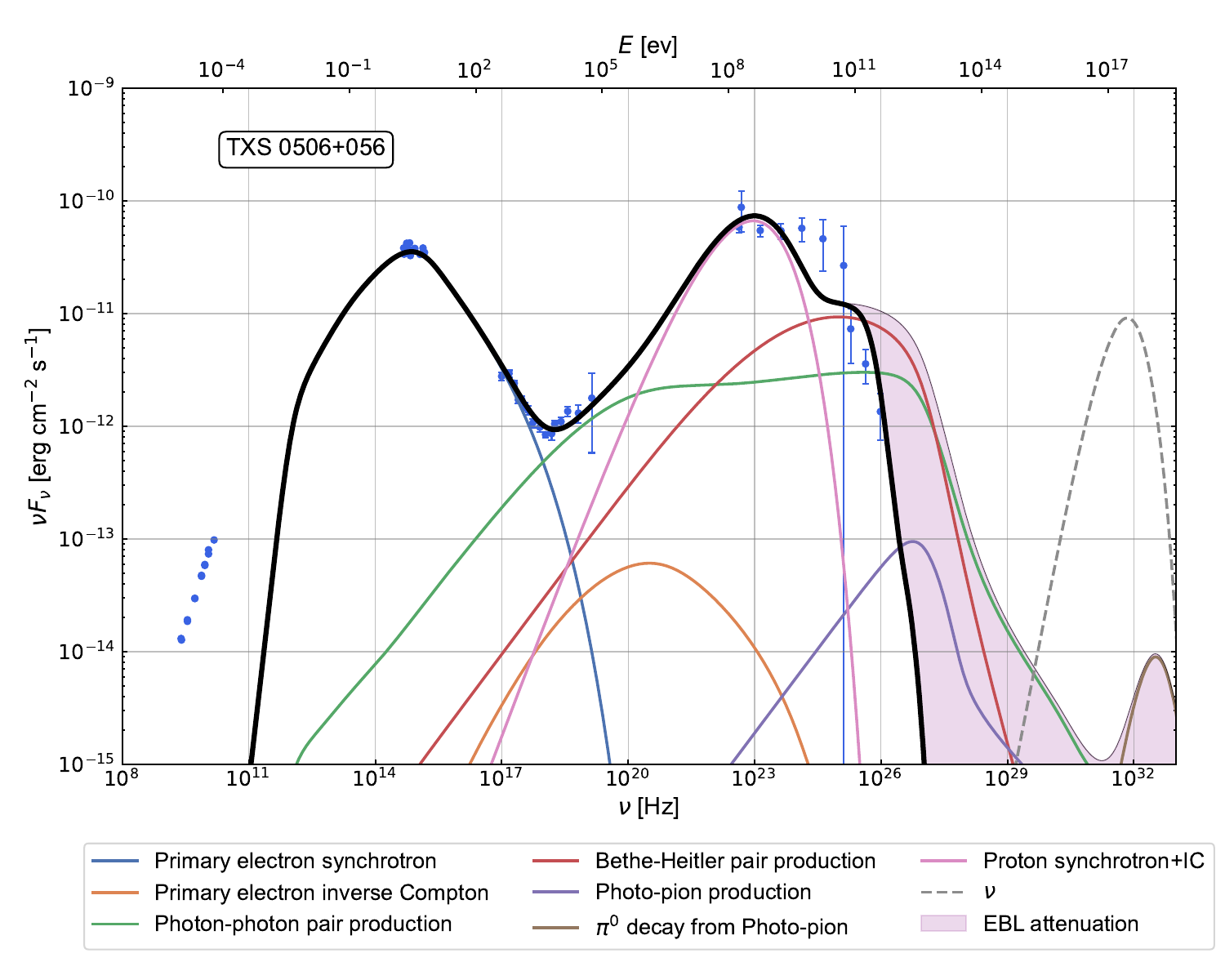}
\caption{
The SED modeling plots for TXS 0506+056 and three NBCs.
The symbols used for the data points and the curves are the same as those adopted in Figure \ref{SED_plot}.
}
\label{SED_neu}
\end{figure}

Four BL Lac objects, TXS~0506+056 and three of our NBCs, lie above the IceCube discovery potential curve.
Within our model, the peak neutrino flux of TXS~0506+056 exceeds the IceCube discovery potential by nearly one order of magnitude, consistent with its confirmed neutrino detection.
The remaining three NBCs — 4FGL~J0045.7+1217, 4FGL~J0123.1+3421, and 4FGL~J0326.2+0225 — only marginally exceed the IceCube sensitivity curve, yet they may still be detectable in neutrinos.
However, none of these three NBCs has been reported as a neutrino emitter at the time of writing.
This discrepancy is expected, as the neutrino fluxes derived in our model represent upper limits, and the actual neutrino emission may be significantly weaker.
The SEDs of these four sources are shown in Figure \ref{SED_neu}, we notice that these sources all show strong neutrino emission that comparable to $\gamma$-ray flux.

At present, TXS~0506+056 remains the only blazar firmly confirmed as a neutrino emitter in the PeV range. 
More recently, the Seyfert~II galaxy NGC~1068 has also been identified as a neutrino source with a significance of $4.2\sigma$ in the TeV band \citep{IceCube2022Sci378}.
This discovery has prompted discussions on whether high-energy neutrinos may predominantly originate from the accretion disk corona rather than from relativistic jets in AGNs.
Consequently, the identification of additional blazars as neutrino emitters would provide a crucial smoking-gun signature for the presence of relativistic hadrons in blazar jets and neutrino emission region in AGNs.

\begin{table}[htbp]
\centering
\caption{Potential detectability of our NBCs by neutrino facilities.}
\label{neu_det}
\begin{tabular}{lcccccc}
\hline
4FGL Name & IceCube & IceCube-Gen2 & KM3NeT & NEON@1700m & NEON@3500m & TRIDENT \\
(1) & (2) & (3) & (4) & (5) & (6)  & (7)    \\
\hline
J0006.3-0620	&		&		&		&		&		&    \\
J0029.0-7044	&		&		&		&		&		& Y  \\
J0045.7+1217	&	Y	&	Y	&	Y	&	Y	&	Y	& Y  \\
J0121.8-3916	&		&		&	Y	&	Y	&	Y	& Y  \\
J0123.1+3421	&	Y	&	Y	&	Y	&	Y	&	Y	& Y  \\
\hline
\end{tabular}
\tablecomments{
column (1) gives the 4FGL name;
column (2) indicates the potential detectability by IceCube, where 'Y' denotes 'Yes'.
column (3) indicates the potential detectability by IceCube-Gen2;
column (4) indicates the potential detectability by KM3NeT;
column (5) indicates the potential detectability by NEON@1700m;
column (6) indicates the potential detectability by NEON@3500m;
column (7) indicates the potential detectability by TRIDENT;
Only five items are displayed, the entire table is available in machine-readable form.}
\end{table}


\section{Conclusion}
\label{sec:con}
In the present work, we extend the study of neutrino blazars following our previous investigation (Paper~I).
We model the broadband SEDs constructed from archival multiwavelength data for 103 sources — comprising 85 BL Lacs, 5 FSRQs, and 13 BCUs — within a hadronic framework.
Our main conclusions can be summarized as follows:
(1) From the broadband SED modeling, we constrain nine free parameters that characterize the physical properties of the emission region and the energy distributions of relativistic particles in jets of these 103 NBCs;
(2) Using a partial correlation analysis, we investigate the relationship between neutrino luminosity and multi-wavelength electromagnetic emission (radio, optical $R$ band, X-ray, and $\gamma$-ray);
The regression results reveal positive correlations between $\log L_{\nu}$ and $\log L_{\rm R}$, $\log L_{\rm X}$, and $\log L_{\gamma}$.
(3) The observed correlation between $\log L_{\rm R}$ and $\log L_{\nu}$ provides potential clues to the neutrino production site.
The neutrino production site may be located either within the jet or associated with the disk/corona system, although it cannot be further constrained with the current analysis;
(4) Our results highlight the MeV band as a key observational window for distinguishing between leptonic and hadronic emission scenarios, as well as for constraining the particle composition of blazar jets;
(5) Based on the maximum neutrino fluxes predicted by our hadronic model, three NBCs are expected to be detectable by IceCube, while up to 22 NBCs may be within the detection reach of KM3NeT in near future.

\acknowledgments
We thank Dr. Rui Xue from Zhengjiang Normal University and Dr. Dahai Yan from Yunnan University for the discussion of blazar neutrino origination.
We thank Dr. Donglian Xu and Qichao Chang from Shanghai Jiao Tong University for sharing the TRIDENT 5$\sigma$ discovery potential curve.
LC acknowledges the support from the National SKA Program of China (2022SKA0120102).
HBX acknowledges the support from the National Natural Science Foundation of China (NSFC 12203034), the Shanghai Science and Technology Fund (22YF1431500), the science research grants from the China Manned Space Project (CMS-CSST-2025-A07), and the Shanghai Municipal Education Commission regarding artificial intelligence empowered research.
LLY acknowledges the support National Natural Science Foundation of China (NSFC) grants 12261141691.
MFG is supported by the National Science Foundation of China (grant 12473019), the Shanghai Pilot Program for Basic Research-Chinese Academy of Science, Shanghai Branch (JCYJ-SHFY-2021-013), the National SKA Program of China (Grant No. 2022SKA0120102), and the China Manned Space Project with No. CMS-CSST-2025-A07.
SHZ acknowledges the support from the National Natural Science Foundation of China (Grant No. 12173026), the National Key Research and Development Program of China (Grant No. 2022YFC2807303), the Shanghai Science and Technology Fund (Grant No. 23010503900), the Program for Professor of Special Appointment (Eastern Scholar) at Shanghai Institutions of Higher Learning and the Shuguang Program (23SG39) of the Shanghai Education Development Foundation and Shanghai Municipal Education Commission.
ZJL acknowledges the support from the Shanghai Science and Technology Foundation Fund under grant No. 20070502400, and the science research grants from the China Manned Space Project.
JHF acknowledges the support from the NSFC U2031201, NSFC 11733001, NSFC 12433004, the National Key Research and Development Program of China (Grant No. 2025YFA1614102), the Scientific and Technological Cooperation Projects (2020–2023) between the People’s Republic of China and the Republic of Bulgaria, the science research grants from the China Manned Space Project with No. CMS-CSST-2021-A06, and the support for Astrophysics Key Subjects of Guangdong Province and Guangzhou City, and the Bulgarian National Science Fund of the Ministry of Education and Science under grants KP-06-H38/4 (2019), KP-06-KITAJ/2 (2020) and KP-06-H68/4 (2022).

\bibliographystyle{aasjournal}
\bibliography{lib_xiao}{}

\end{document}